\documentclass[letterpaper]{jpconf}

\begin{document}

\newcommand{\mic}{$\mu$m\  } 

\title{Upper Limits to Fluxes of Neutrinos and Gamma-Rays from Starburst Galaxies}

\author{F.W. Stecker$^1$}

\address{$^1$ Astrophysics Science Division, NASA Goddard Space Flight Center, Greenbelt, MD, USA}\

\newcommand{\solar}{\ifmmode_{\mathord\odot}\else$_{\mathord\odot}$\fi}
\newcommand{\gray}{ $\gamma$-ray}
\newcommand{\grays}{ $\gamma$-rays}
\newcommand{\apj}{ Astrophys. J.}
\newcommand{\prl}{ Phys. Rev. Letters}

\begin{abstract}

Loeb and Waxman have argued that high energy neutrinos from the
decay of pions produced in interactions of cosmic rays with  
interstellar gas in starburst galaxies would be produced with
a large enough flux to be observable. Here we obtain an upper 
limit to the diffuse neutrino
flux from starburst galaxies which is a factor of $\sim$5 lower than
the flux which they predict. 
Compared with predicted fluxes from other extragalactic high energy
neutrino sources, starburst neutrinos with $\sim$ PeV energies
would have a flux considerably below that predicted for AGN models.
We also estimate an upper limit for the diffuse GeV $\gamma$-ray flux from
starbust galaxies to be $\cal{O}$$(10^{-2})$ of the observed $\gamma$-ray
background, much less than the component from unresolved blazars.
\end{abstract}

\section{Introduction}

Interactions of cosmic-ray nuclei with interstellar gas nuclei 
in our galaxy produce $\pi^0$'s which 
decay to produce most of the galactic $\gamma$-rays above 0.1 GeV 
~\cite{st77}; the decay of the $\pi^+$'s produced 
yields galactic cosmic-ray neutrinos~\cite{st79}.

The distribution of high
energy $\gamma$-rays in our Galaxy is related to the distribution
of molecular clouds and very young hot high-mass stars in OB 
associations which are short-lived and explode into 
supernovae~\cite{st75,st76}.
This association between supernovae which are likely to produce 
cosmic rays and dense regions of molecular gas led to the scenario
where interactions between the gas and cosmic rays 
produces the galactic $\gamma$-rays {\it via} the decay of the $\pi^0$.
A natural implication then would be that
starburst galaxies, which are undergoing a phase of extremely active star 
formation would be likely sources of high energy $\gamma$-rays 
\cite{to04,to05}. 

Loeb and Waxman (LW) ~\cite{lw06} have suggested that such hadronic 
processes in starbust galaxies can produce, {\it in toto},
a large enough background of diffuse high energy neutrinos to
be observable with a very large neutrino detector
such as {\it Icecube}~\cite{ha06}. LW then argue 
that radio observations of starburst galaxies imply a {\it lower limit} on
the cumulative extragalactic neutrino flux from starburst galaxies 
which is within the sensitivity range of {\it Icecube}. We derive here an
{\it upper limit} to the cumulative high energy neutrino flux from
starburst galaxies which is significantly lower than the LW ``lower limit''. 
The diffuse flux of GeV $\gamma$-rays from the same processes
is found to be $\cal{O}$$(10^{-2})$ of the observed $\gamma$-ray
background, much less than the component from unresolved blazars and
more than an order of magnitude below a recent estimate~\cite{th06}.

\section{Radio Emission and the Neutrino Flux from Starbust Galaxies}

LW start with the observed synchrotron
emission from starburst galaxies which is produced by relativistic
electrons in these sources~\cite{yu01}. They then make the assumptions
that (1) the presence of relativistic electrons in these sources implies 
the presence of relativistic protons, (2) the protons lose essentially
all of their energy to pion production, and (3), a lower limit to the
energy loss rate of the protons can then be obtained from the synchrotron
radio flux by assuming that all of the electrons (and positrons) which
are radiating are from pion decay. 

Assumption (1) is a reasonable one which is supported by observations
of cosmic rays in our own Milky Way galaxy. Assumption (3), {\it viz.}, the
``lower limit'' assumption depends on assumption (2). However,
assumptions (2) and (3) can be questioned because (a) the synchrotron radiating
electrons may be largely accelerated primaries rather than secondaries
related to pion production and decay, as is the case in our own Galaxy,
and (b) the conditions in starburst
galaxies are significantly different from those in our own galaxy.
In particular, starburst galaxies exhibit strong ``superwinds''~\cite{ho03}. 
Such winds have significant dynamical effects and may disrupt
magnetic fields and drive
protons out of these galaxies before they can lose all of their energy
by interacting with interstellar gas nuclei to produce pions~\cite{ar}.
In contrast, assumption (2) of LW assumes full trapping of relativistic
nuclei in the disks of starburst galaxies to the point where they only
lose energy in hadronic interactions with gas atoms. This is in stark 
contrast to the situation in our own Galaxy (see footnote 1) and is
also contradicted by recent observations of high redshift galaxies~\cite{vl}.

These caveats call into serious question the argument that the
radio data can provide a lower limit on the cumulative
diffuse flux of neutrinos from starburst galaxies. But here we
consider that assumptions (1)-(3) can be used for 
obtaining an analytic {\it upper limit} for such a flux. We also
will accept the other estimates which lead to the ratio of injected power
of protons to electrons at a fixed particle energy, $\eta_{p/e} \simeq 6$
and a neutrino luminosity which is then related to the local radio 
luminosity density by 

\begin{equation}
E_{\nu}^2\Phi_{\nu} (E_{\nu} = 1 GeV) \simeq (ct_{H}/4\pi)\zeta
[4f(dL_{f}/dV)]_{f = 1.4 GHz}
\end{equation}

\noindent where $t_{H}$ is the age of the universe
and $\zeta = 3$ is an evolution factor which takes account of the fact
that starburst galaxies were more numerous in the past~\cite{lw06}.
 LW take the local energy production rate per unit 
volume at a frequency f = 1.4 GHz to be $\simeq 10^{28.5}$ W Mpc$^{-3}$.
Let us reexamine this value for $f(dL_{f}/dV)]_{f = 1.4 GHz}$.

The local 1.4 GHz energy production rate has been derived by LW
by making use of the correlation bewteen GHz and far
infrared (FIR) emission in galaxies given by 
Yun, Reddy and Condon (YRC)~\cite{yu01}. YRC use the data on
{\it IRAS} galaxies to derive the local infrared luminosity density at 
60 $\mu$m to be $2.6 \times 10^7 L_{\solar}$ Mpc$^{-3}$. 
This {\it total}
power density is then used by LW to obtain the 1.4 GHz power density {\it via} 
the relation given by YRC.
The key difference between the result derived here and that obtained
by LW is in chosing how to interpret the paper of YRC.
YRC state that less than 10\% of the local FIR luminosity density 
is contributed by luminous IR galaxies with $L_{FIR} > 10^{11} L_{\solar}$;
this is the component which includes the starburst galaxies. (Figure
11 of YRC yields an estimate of $\sim$ 6\%.)
We therefore take the local contribution from starburst galaxies alone
at 60$\mu$m to be $< 2.6 \times 10^6L_{\solar}$Mpc$^{-3}$. Consequently, 
the component of the local radio luminosity density from starburst galaxies is 
$\Phi_{1.4  GHz} <  10^{27.5}$ W  
Mpc$^{-3}$\footnote{One might ask ``Why not consider
``normal'' galaxies with lower FIR luminosities and add them in to
estimate a higher neutrino flux?'' However, in normal
galaxies like ours, cosmic rays lose only a small fraction of their 
energy {\it via} hadronic interactions, contrary to
assumption (2). Also galactic PeV cosmic rays have a much steeper spectrum 
than the $E^{-2}$ assumed by LW.}

There is a higher relative fraction of the
energy input from the higher relative number of starburst galaxies at higher 
redshifts. The fraction of 
the FIR background, $\kappa(\Delta z)$ contributed by galaxies in different 
redshift ranges, $\Delta z$, is obtained from Ref.~\cite{la05}. We 
multiply $\kappa(\Delta z$) by the fraction of the FIR background contributed 
by starburst galaxies in different redshift ranges, $\xi(\Delta z)$, 
to estimate the mean fraction of the total FIR background contributed by 
starburst galaxies. Estimates for $\kappa$ and $\xi$ are shown in Table 1.
Using the results from Table 1, we estimate that 23\% of the observed
FIR background integrated over redshift is from starburst galaxies.

\begin{table*}[ht]
\vspace{0.3cm}
\centerline{Table 1: Relative contributions to the $\nu$ Starburst Galaxy Flux (see text).} 
\vspace{0.2cm} 
\begin{center}
\begin{tabular}{cccc}  \hline \hline

Redshift Range ($\Delta z$) & $\kappa(\Delta z)$~\cite{la05} & $\xi(\Delta z)$ & Reference for $\xi$ \\
\hline
0 to 0.2 & 10\% & $<$ 10\% & \cite{yu01} \\
0.2 to 1.2 & 68\% & $\sim$ 13\% & \cite{la05}
 \\
$>$1.2  & 22\% & $\sim$ 60\% &  \cite{er06} \\

\hline

\end{tabular}
\end{center}
\vspace{0.6cm}
\end{table*}

\section{Observability of High Energy Neutrinos from Starburst Galaxies}

The upper limit on the radio flux from starburst galaxies obtained above can 
be used to obtain an upper limit on the neutrino flux from starburst
galaxies by using equation (1) of LW. One then finds that the  
neutrino background
energy  flux from starburst  galaxies  $<  2 \times  10^{-8}$ 
GeV cm$^{-2}$ s$^{-1}$ sr$^{-1}$ 
Such a flux  would be undetectable above the atmospheric background
neutrino flux, even  if  equation (1)  is  assumed to  be 
valid  when extrapolated to 300 TeV and  even granting all of the 
assumptions made by LW.\footnote{Even
if we make a second extreme assumption that 100\% of the IR galaxies
at redshifts greater than 1.2 are starburst galaxies, we would still
predict a neutrino flux $< 3 \times 10^{-8}$ GeV cm$^{-2}$ s$^{-1}$ sr$^{-1}$.
If we use the fast evolution model of Ref.~\cite{sms}, we would
obtain a similar upper limit.}

\begin{table*}[ht]
\centerline{Table 2: Neutrino Energy Fluxes (GeV cm$^{-2}$ s$^{-1}$ sr$^{-1}$)}
\begin{center}
\begin{tabular}{ccccc}  \hline \hline

$\nu$ Source &  $E^2\Phi(10 \rm TeV)$ &  $E^2\Phi(100\rm TeV)$ & $E^2\Phi(1 \rm PeV)$ & Reference \\
\hline 
Atm: AMANDA-II & $2 \times 10^{-6}$ & $7 \times 10^{-8}$ & $<3 \times 10^{-9}$ & \cite{bo06} \\
Atm (Vertical) & $7 \times 10^{-7}$ & $\sim 2 \times 10^{-9}$  & --- & \cite{ga05} \\ 
AMANDA-II Diff.Lim. & $9 \times 10^{-8}$ & $9 \times 10^{-8}$ & $9 \times 10^{-8}$ & \cite{hi06} \\ \hline

Starburst Galaxies & $< 2 \times 10^{-8}$ & $< 2 \times 10^{-8}$ & $< 2 \times 10^{-8}$ & This paper \\ 
AGN Cores &   $5 \times 10^{-10}$ & $10^{-8}$ & $10^{-7}$  &\cite{st05} \\ 
AGN & $3 \times 10^{-9}$ & $3 \times 10^{-8}$ & $2 \times 10^{-7}$ & \cite{mpr} \\ 
GRB & $5 \times 10^{-10}$ & $3 \times 10^{-9}$ & $3 \times 10^{-9}$ & \cite{wb} \\ \hline
Icecube Sensitivity & --- & $4 \times 10^{-9}$ & $4 \times 10^{-9}$ & \cite{ri05} \\
\hline
\end{tabular}
\end{center}
\end{table*}

Table 2 shows a comparison of the upper limit on the flux from
starburst galaxies given here with the atmospheric neutrino flux and
with approximate model predictions of neutrino fluxes $\gamma$-ray bursts
(GRB) and active galactic nuclei (AGN) along with detector array
sensitivities. It can be seen from this table that at 100 TeV none of the
extragalactic sources proposed will dominate over the atmospheric
foreground. The present upper limit on the diffuse neutrino energy 
flux below 1 PeV from AMANDA-II is 
$\sim8.8 \times 10^{-8}$ GeV cm$^{-2}$ s$^{-1}$ sr$^{-1}$ 
in the 10 TeV 
to 1 PeV energy range~\cite{hi06}. The full {\it Icecube} detector
array is expected to push down to a sensitivity of 
$\sim 10^{-9}$ GeV cm$^{-2}$ s$^{-1}$ sr$^{-1}$ in the
energy range 100 TeV $< E_{\nu} <$ 100 PeV after several years of
observation.
Under the extreme assumption that the primary cosmic ray spectra 
in all starburst galaxies are as hard as $E^{-2}$ up to energies 
$\cal{O}$(10 PeV), PeV neutrinos from starburst galaxies may be 
detectable just above the projected sensitivity of {\it Icecube}.
However, as can be seen from Table 2, above 1 PeV the 
AGN models predict fluxes which will be significantly larger than  
the the atmospheric foreground (expected to be $< 3 \times 10^{-9})$ 
GeV cm$^{-2}$ s$^{-1}$ sr$^{-1}$ at 1 PeV), as well as the fluxes from 
$\gamma$-ray bursts (GRB) and starburst galaxies.

\section{Observability of Diffuse $\gamma$-rays from Starburst Galaxies}

In a follow-up paper to LW, Thompson {\it et al.}~\cite{th06}
have estimated the contribution of $\pi^{0}$-decay $\gamma$-rays from
starburst galaxies to the 
observed $\gamma$-ray background in the GeV energy range. Using an $E^{-2}$ 
primary spectrum they get estimates of $E^2\Phi(E)$ fluxes for both 
$\gamma$-rays and neutrinos of $3 \times 10^{-7}$ GeV 
cm$^{-2}$ s$^{-1}$ sr$^{-1}$.

Our estimated starburst galaxy $\gamma$-ray flux for the
same $E^{-2}$ primary spectrum assumed in Ref.~\cite{th06} is 
$\sim 2 \times 10^{-8}$ GeV 
cm$^{-2}$ s$^{-1}$ sr$^{-1}$ for $\gamma$-ray energies less than $\sim 10$ GeV.
This is $\cal{O}$$(10^{-2})$ of the observed flux of $\sim 1.4 \times
10^{-6}$ GeV cm$^{-2}$ s$^{-1}$ sr$^{-1}$ determined by the {\it EGRET}
group~\cite{sr98} and is therefore unobservable. 
Above $\sim 10$ GeV the background spectrum will
steepen owing to absorption from pair production interactions with
the extragalactic ultraviolet background radiation~\cite{ss96}.
Stecker and Salamon have shown 
that the bulk of the observed background can be
produced by unresolved blazars~\cite{ss96}.  
Note, almost all of the observed extragalactic GeV $\gamma$-ray sources 
are blazars; no starburst galaxies have been observed.
\section*{Acknowlegdments}

The author would like to thank Casey Papovich of the {\it Spitzer} team
and Dawn Erb for their comments on extragalactic infrared emission and 
Eliot Quataert for his comments. This work was supported by NASA Grant 
ATP03-0000-0057.

\end{document}